\documentclass[12pt,preprint]{aastex}

\usepackage[utf8]{inputenc}
\usepackage{natbib}
\usepackage{epstopdf}
\usepackage{graphicx}

\newcommand\tstrut{\rule{0pt}{2.5ex}}

\slugcomment{}

\shorttitle{Activation of Main Belt Comets}
\shortauthors{Haghighipour et al.}

\begin{document}

\title{Triggering Sublimation-Driven Activity of Main Belt Comets}

\author{N.\ Haghighipour
}
\affil{Institute for Astronomy, University of Hawaii-Manoa, Honolulu, HI, 96825, USA }
\email{nader@ifa.hawaii.edu}

\author{T.\ I.\ Maindl}
\affil{Department of Astrophysics, University of Vienna,  T\"{u}rkenschanzstrasse 17, A-1180 Vienna, Austria}

\author{C.\ Sch\"{a}fer}
\affil{Institut f\"{u}r Astronomie und Astrophysik, Eberhard Karls 
Universit\"{a}t T\"{u}bingen, Auf der Morgenstelle 10, 72076 T\"{u}bingen, Germany}

\author{R.\ Speith}
\affil{Physikalisches Institut, Eberhard Karls Universit\"{a}t T\"{u}bingen, 
Auf der Morgenstelle 14, 72076 T\"{u}bingen, Germany}

\and

\author{R.\ Dvorak}
\affil{Department of Astrophysics, University of Vienna,  T\"{u}rkenschanzstrasse 17, A-1180 Vienna, Austria}

\begin{abstract}
It has been suggested that the comet-like activity of Main Belt Comets are due to the sublimation of sub-surface 
water-ice that has been exposed as a result of their surfaces being impacted by m-sized bodies. We have examined 
the viability of this scenario by simulating impacts between m-sized and km-sized objects using a smooth particle 
hydrodynamics approach. Simulations have been carried out for different values of the impact velocity and impact 
angle as well as different target material and water-mass fraction. Results indicate that for the range of impact 
velocities corresponding to those in the asteroid belt, the depth of an impact crater is slightly larger than 10 m 
suggesting that if the activation of MBCs is due to the sublimation of sub-surface water-ice, this ice has to exist 
no deeper than a few meters from the surface. Results also show that ice-exposure occurs in the bottom and on the 
interior surface of impact craters as well as the surface of the target where some of the ejected icy inclusions 
are re-accreted. While our results demonstrate that the impact scenario is indeed a viable mechanism to expose ice and 
trigger the activity of MBCs, they also indicate that the activity of the current MBCs is likely due to ice 
sublimation from multiple impact sites and/or the water contents of these objects (and other asteroids in the outer 
asteroid belt) is larger than the 5\% that is traditionally considered in models of terrestrial 
planet formation providing more ice for sublimation.
We present details of our simulations and discuss their results and implications.

\end{abstract}


\section{Introduction}
The detection of comet-like activities in several asteroids during the past decade has generated much
interest in the origin of these objects and the mechanism of their activation. Dubbed Main Belt Comets
(MBCs), these bodies show dynamical behaviors characteristic of main belt asteroids (e.g., their orbits are 
entirely in the asteroid belt, some belong to asteroid families, and similar to other main belt asteroids, 
their Tisserand parameters with respect to Jupiter\footnote{For a small object, such as an asteroid, that 
is subject to the gravitational attraction of a central star and the perturbation of a planetary 
body P, the quantity ${a_{\rm P}}/a \, + \,2 {[(1-e^2)\,a/{a_{\rm P}}]^{1/2}}\, \cos i$ is defined as its 
Tisserand number. In this formula, $a$ is the semimajor axis of the object with respect to the star, $e$ is its 
orbital eccentricity, $i$ is its orbital inclination, and $a_{\rm P}$ is the semimajor axis of the planet. 
In the solar system, the Tisserand number of a small body with respect to Jupiter has, traditionally, been used 
as a determinant of the cometary or asteroidal nature of its orbit. In general, the Tisserand numbers of comets 
with respect to Jupiter are smaller than 3, whereas those of asteroids are mostly larger. See also \citet{Hsieh16} 
for the extent of the viability of using Tisserand number with respect to Jupiter as an orbital determinant.} 
are larger than 3), whereas unlike 
most asteroids, they carry tails similar to those of comets. Much of interest in these objects is due to the 
implication that their comet-like tails and activity are driven by the sublimation of volatile material, 
presumably water-ice. The latter strongly argues in favor of the idea that water-bearing asteroids, with origins 
at the outer part of the asteroid belt have been the source of the majority of Earth's water.

At the time of the writing of this article, 8 unambiguous MBCs\footnote{We call an MBC unambiguous if its
activation can only be explained by sublimation of volatiles that have been surfaced as a result of its collision with 
a small impactor.} were known: 133P/(7968) Elst-Pizzaro, 176P/(118401) LINEAR, and 238P/READ (P/2005 U1) \citep{Hsieh06}, 
259P/Garradd(P/2008 R1) \citep{Jewitt09}, P/2010 R2 (La Sagra) \citep{Hsieh12a},
288P/(300163) P/2006 ${\rm VW}_{139}$ \citep{Hsieh12b}, P/2012 T1 (PANSTARRS) \citep{Hsieh13}, and 313P/Gibbs (P/2014 S4) \citep{Hsieh15}.
Table 1 and Figure 1 show the sizes and orbital properties of these objects as well as their locations in the asteroid belt. 
Among these MBCs, 133P/(7968) Elst-Pizzaro is a member of the 10 Myr old Beagle family\footnote{Beagle family is a 
sub-family of the larger Themis family \citep{Nesvorny08}.}\citep{Nesvorny08}, and P/2012 T1 (PANSTARRS) and 313P/Gibbs (P/2014 S4) 
are associated with the $\sim 155$ Myr Lixiaohua family \citep{Hsieh13,Hsieh15}. 

Studies of the dynamical evolution of 133P/(7968) Elst-Pizzaro, 176P/(118401) LINEAR, and 238P/READ (P/2005 U1) have
indicated that while the orbits of 133P/(7968) Elst-Pizzaro and 176P/(118401) LINEAR are stable for 1 Gyr,
238P/READ (P/2005 U1) becomes unstable in
about 20 Myr \citep{Hagh09}. Given the association of 133P/(7968) Elst-Pizzaro with the Beagle family, and the proximity of
176P/(118401) LINEAR, and 238P/READ (P/2005 U1) to the Themis family (Figure 2), these results suggested that
133P/(7968) Elst-Pizzaro, 176P/(118401) LINEAR, and 238P/READ (P/2005 U1)
are most likely natives to the asteroid belt and formed in-situ through the breakage of larger asteroidal bodies
\citep{Hagh09,Hagh10}. 
The orbital instability of 238P/READ (P/2005 U1) can then be attributed to the interaction of this MBC with Jupiter through the 2:1 
mean-motion resonance. As shown by \citet{Hagh09}, interactions with mean-motion resonances  
could have affected the dynamics of other MBCs as well causing them to be scattered
into orbits away from their birth places. As suggested by this author, the latter implies that, 
while asteroid families and associations in the outer part of the asteroid belt present promising regions 
to search for MBCs, stand-alone MBCs can also exist and may be found throughout the asteroid belt. 
The subsequent detection of MBC 259P/(Garradd)(P/2008 R1) by \citet{Jewitt09} supported this prediction.

It is important to note that the dynamical analyses presented in \citet{Hagh09} address only the origin of
MBCs and not that of their larger parent bodies. While there is no reason to believe that the parent bodies of MBCs 
could not be native
to the asteroid belt, it is also plausible to assume that these objects might have formed in the outer regions of the
solar system and were scattered into the asteroid belt during the post-formation evolution of giant planets.
Results of the dynamical analysis by \citet{Hagh09}, as well as those carried out for other 
MBCs \citep{Jewitt09,Hsieh12a,Hsieh12b,Hsieh13,Hsieh15}, however, suggest that current km-sized MBCs 
most probably formed in the asteroid belt as a result 
of multiple breakage and fragmentation of their parent and grandparent bodies.
We refer the reader to \citet{Hsieh16} for more details.

An interesting characteristics of the currently known MBCs is their small sizes. As shown in Table 1,
these objects are smaller than 5 km in size. This small sizes of MBCs has strong implications for the 
history and mechanism of the activation of these objects. \citet{Hsieh04} and \citet{Hsieh06} studied 
the mass loss and activation of 133P/(7968) Elst-Pizzaro and showed that, considering an albedo of 0.05 
and temperature of 150 K, this MBC reduces size at a rate of 1 m/yr. 
Given that the diameter of 133P/(7968) Elst-Pizzaro is slightly smaller than 4 km, such a rapid rate of size-reduction 
strongly suggests that the activation of this MBC has been recent. Taking into account the recurrent activity of
133P/(7968) Elst-Pizzaro in 1996, 2002 and 2004, and also observing an increase followed by a decrease in the activity of this MBC 
during its 2004 perihelion passage, \citet{Hsieh04} and \citet{Hsieh06} suggested that the activity of 133P/(7968) Elst-Pizzaro 
could not be impact-generated and is probably due to the sublimation of some volatile material, presumably water-ice 
(an assumption consistent with the 
location of this MBC in the outer regions of the asteroid belt). These authors suggested that 133P/(7968) Elst-Pizzaro 
was impacted by a 1-10 m-sized body which excavated its sub-surface ice causing ice to sublimate and create a thin 
atmosphere and a tail when the orbit of this MBC brings it close to the Sun.

The degree to which the impact of a meter-sized body with a km-sized MBC can excavate sub-surface ice and
trigger MBC's activity depends on whether the resulted impact crater will be deep enough to reach the underlying
water-ice. The depth of the crater, itself, depends on the impact velocity of the m-sized impactor
and the material strength of the MBC. The purpose of this study is to examine the viability of the 
above-mentioned scenario by simulating the collision
of a m-sized body with a km-sized object for different impact velocities and different materials of the
impacted body. Specifically, we aim to answer the following three questions:
\begin{itemize}
\item What is the maximum depth of an impact crater in a collision between a meter-sized object 
and a km-sized body for typical impact velocities in the asteroid belt?
\item Can such impacts result in the exposure of sub-surface water-ice? In other words, can these impacts
differentiate between models of ice-longevity in asteroidal bodies?
\item Does the activity of an MBC originate solely from the sublimation of the exposed ice in its crater, 
or the re-accreted ice will also contribute (and to what extent)?
\end{itemize}
\noindent 
To answer these questions, we carried out a large number of 3D smooth particle hydrodynamics (SPH) 
simulations of a collision between a m-sized body and a km-sized object. 
Simulations were carried out for different values of impact velocities and 
different values of the material strength and water content of the impacted body. 

The outline of this paper is as follows.
We begin by describing the details of our model and computational method in section 2. 
Section 3 has to do with the presentation of our results and their implications for the exposure of 
sub-surface water-ice and models of ice-longevity. We also present in this section a detailed study 
of the area of ice exposure and its connection to the activation of MBCs. In section 4, we conclude 
this study by presenting a summary of our findings and their limitations. We also discuss the implications 
of our results for the water content of asteroidal bodies and their contribution to the Earth's water.

\section{SPH simulations and initial set up}

As mentioned in the introduction, the main purpose of this study is to examine the viability of the impact
scenario as a mechanism for exposing water-ice and to differentiate between models of ice-survival in 
the asteroid belt. For this purpose, and because the depth of an impact crater strongly depends on the 
material strength of the impacted body, we assume a large range of material strength and consider targets made of 
a mixture of ice with very hard and very soft materials. 
For high strength material, we consider basalt and for low strength
material, we use tuff. For the sake of comparison, we also consider targets with 100\% ice.
Although actual MBCs are most likely carbonaceous chondrites, such a choice of target material will
allow us to portray a general picture of the impact scenario, and determine the ranges of size and depth of impact craters
for a large range of water content and material strength.

We use a 3D SPH code developed by \citet{Schafer07,Schafer16} and \citet{Maindl13} to simulate an impact between two bodies.
Our code solves the continuity equation and the equation of the conservation of momentum in continuum mechanics.
The continuum of the solid body is discretized into mass packages known as SPH particles. The locations of these
particles are the sampling points of our numerical method. The SPH particles move as point masses according to the
equation of motion. Each particle carries all physical properties (e.g.,  mass, momentum, energy) of the part of the solid
body that it represents. Depending on the type of material, particles may have different material parameters such
as bulk and shear modulus, yield strength or different activation thresholds for the development of cracks.

Our code also includes material strength and implements a full elasto-plastic continuum mechanics 
model \citep[see, e.g.,][]{Maindl13,Maindl14}. To model any specific material, we use the Tillotson equation of state 
\citep{Tillotson62,Melosh96} with 
material parameters as listed in Table 2. To treat fracture and brittle failure, we use the Grady-Kipp 
fragmentation prescription \citep{Grady93,Benz94,Benz99}. This prescription is based on flaws that are distributed in 
the material following a Weibull distribution with material-dependent parameters. 

When considering basalt or when the target is pure ice, we use the values given by
\citet{Maindl13} for rock and ice parameters in the equation of state and the Weibull distribution. 
These values are based on measurements by \citet{Nakamura07} and can be found in their Table 1.
Basaltic rock, represented by the parameters of the basalt equation of state, as given by \citet{Melosh96}, 
is widely used as the material of rocky bodies from cm-sized up to the mantles of large asteroids such as Vesta 
or Ceres \citep{Melosh97,Agnor04,Nakamura09,Michel09}. We use slightly modified parameters  
similar to those used by \citet{Benz99}, and
apply a tensorial correction as suggested by \citet{Schafer07,Schafer16} to ensure first-order consistency. 
When simulating collisions, we implement our damage model using Weibull parameters $m$ and $k$. 
For basalt, we use $m=16$  and $k={10^{61}} ({\rm {m^{-3}}})$ \citep{Nakamura07} and for water-ice, 
we use $m=9.1$ and  $k={10^{46}} ({\rm {m^{-3}}})$ \citep{Lange84}. 

When considering tuff, because there are no data on the flaw distribution in this material, and also because tuff is not as 
brittle as ice or basalt, we do not model fractures. Instead, we assume ductile deformation and fragmentation.

To simulate collisions, 
we resolve the combined system of the impactor and target into approximately 500,000 SPH particles. 
Following \citet{Maindl14} and because the impact timescales are very short compared to the time of the
influence of the gravitational force of the target body (the collision velocities are in the order of km/s 
whereas the MBCs' surface escape 
velocities are less than a few m/s, see Table 1), we simulate collisions without self-gravity. 
To analyze the evolution of the system during each impact, we take 250--500 snapshots every 0.4\,ms. 
In between the snapshots, time integration is continued with an adaptive step-size.

\section{Impact simulations}

We simulated collisions between a m-sized impactor and a km-sized body with the target being a mix of basalt and ice,
tuff and ice, or pure water-ice.
When using basalt and tuff, we considered the water-mass fraction of the target to be 0\%, 5\% 
\citep[following the conventional consideration of the water-mass fraction of asteroids in models of terrestrial 
planet formation as in e.g.,][]{Raymond04,Raymond09,Izidoro13,Izidoro14,Hagh16}, and 50\%.
Because the size and depth of an impact crater varies with the energy imparted to the target, we carried out simulations 
for different values of impact velocity and impact angle. 

The fact that the activation of MBCs is recent indicates 
that irrespective of the origin of these objects, 
collisions between impactors and MBCs must have happened when the projectile and target were both in the asteroid belt.
As shown by \citet{Bottke94}, the mean impact velocity of bodies in the
asteroid belt is $\sim 5.3$\,km ${\rm s}^{-1}$ with the most probable value being around 
4.4\,km ${\rm s }^{-1}$  for bodies larger than 50\,km.
This indicates that the collision velocity distribution, while having a long high-velocity tail is 
heavier towards smaller velocities. We assume an abundance of objects significantly smaller than 
$50$\,km on similar orbits (i.e., comparable semi-major axes, eccentricities 
and inclinations) with $e\lesssim 0.25\/$, inside the asteroid belt. Given their similar orbits and in the 
absence of significant mutual gravitational interaction, collisions between these bodies will occur with velocities 
much smaller than several km ${\rm s}^{-1}$. Combining this assumption with the 
finding of Bottke et al. (1994), we therefore considered impact velocities in the range of 
0.5 km ${\rm s}^{-1}$ to 5.3 km ${\rm s}^{-1}$, and added an additional tail to our velocity 
distribution towards smaller values extending down to 100 m ${\rm s}^{-1}$. We also chose the impact 
angle ($\alpha\/$) to vary in the range of 0 (head-on collision) to $60^\circ\/$.

Figure 3 shows snapshots of a sample of our simulations. The target is made of basalt and has 5\% water in mass.
The impact velocity is 4.4\,km ${\rm s}^{-1}$ and the impact angles are 0 and $30^\circ$. To reduce the amount 
of unnecessary computations, we limited the region of simulation in the target to a volume of 
30 m $\times$ 30 m $\times$ 30 m. Figure 4 shows the final state of head-on  $(\alpha=0)$ impacts
for tuff and basaltic targets with water-mass fractions of 0\%, 5\% and 50\%. 
As shown in these figures, ice particles are exposed on the bottom and walls of the crater, and the depth and
morphology of the impact crater vary with the type of the target material and its water content. In the next section,
we present an analysis of these results and discuss their implications 
for the models of ice-longevity in asteroids and the activation of MBCs.

Figures 3 and 4 also show that the debris generated after the impact contains both ice and rock. 
While some of these ice particles and rock debris leave the entire impacted body, 
some are re-accreted and settle on the surface of the target. We discuss this process and its contribution to the
activation of MBCs later in this section.
We note that we consider the entire energy of impact to be imparted to the target 
and do not consider evaporation of ice particles due to the heat of the impact. We allow shocks to propagate throughout
the target and analyze the target's response using our damage model as described in the previous section.
Movies of some of our impact simulations can be downloaded from the electronic supplementary material.

\subsection{Calculation and analysis of crater depth}

Theoretically, the depth of a crater is given by the distance of its deepest point to the top of the crater. 
In the context of our SPH simulations, this implies 
that to determine the depth of a crater, we need to identify the crater's deepest SPH particle.
However, because in our simulations, impact velocities are high, craters form in a small fraction of a
second after an impact. This, combined with the negligible gravity of the target results in the appearance
of spurious SPH particles. The latter makes the
calculation of the depth of a crater based on the locations of individual SPH particles, unreliable. 

To avoid this complication, we measure the depth of a crater using an indirect method.
We begin by considering a cross section of the crater in the direction of the vertical component of the impact velocity
and passing through the crater's plane of symmetry. In this cross section, we fit an ellipse to the crater. 
To ensure that our measurement of the depth of the crater includes a sufficient number of SPH particles at the bottom 
of the crater, we also consider a width for this fitted ellipse in the direction perpendicular to the above cross section
(i.e., in the direction parallel to the horizontal component of the impact velocity). This width  
is chosen such that while the fitted elliptical band
contains sufficiently large number of particles to encompass the bottom of the crater, its particle number will also be small  
enough to avoid obscuring the crater's shape. As mentioned above, impact craters form in a fraction of a second. 
To ensure that prior to the fitting, post-impact variations in the shape and depth of the crater 
have become negligibly small, we allow a time lapse of $\Delta t$. This quantity is chosen such that 
it is as large as necessary to get the final shape of the crater and as small as possible to avoid the shock waves
of the impact to hit the boundary before the crater forms.
Next, while ignoring single, scattered SPH particles, we determine the location of the bottom of the elliptical band
with deepest penetration into the target. We define the crater depth to be the distance between this deepest 
point to the surface of the target. Figure 5 illustrates this methodology for an
impact velocity of 4.4\,km ${\rm s}^{-1}$  at a $30^\circ$ angle onto a basaltic target with 5\% water-ice.
As shown here, the crater has a depth of 3.39 m.
Uncertainties in calculating the depth of a crater in this method depend on the smoothing length of the simulations 
which varies by the size of the target in each individual impact. 
We estimate the errors to be in the range of 0.2 -- 0.5 m.

\subsection{Implications for models of ice-longevity in asteroids}

In order for water-ice to be exposed, the impact crater has
to be deep enough to reach the MBC's internal ice. While the depth of an impact crater strongly depends on the
impact velocity, the depth where the ice is buried depends on the physical and dynamical properties of the MBC. 
As suggested by \citet{Schorghofer08}, in asteroids with mean surface temperatures smaller than 145 K (e.g., the MBC 
133P/(7968) Elst-Pizarro), a layer of only a few tens of centimeter of dust would be able to prevent the ice from 
sublimating and would allow the asteroid to maintain its internal ice for the age of the solar system. 
According to this model, an impact crater of only a few meters in depth 
would be sufficient to expose sub-surface water-ice.
A different model by \citet{Prialnik09}, however, makes drastically different predictions. These authors suggest that 
an MBC such as 133P/(7968) Elst-Pizarro can maintain only crystalline water-ice and only in depths ranging from 50 m to 150 m.

We use the results of our impact simulations to differentiate between these two models.
Figure 6 shows crater depths in all our simulations in terms of the vertical component of 
the impact velocity. 
As expected, the depth of a crater strongly depends on the material strength of the target.
For a given value of the vertical component of the impact velocity, the crater depth increases 
by varying target material from basalt to tuff, and more importantly, by increasing target's
water-mass fraction.
An interesting results shown in figure 6 is that even when the target is considered to be pure ice, the 
crater depth does not exceed 12 m. For the conventional case of 5\% water-mass fraction, the depth of the 
crater stays below 6 m in all simulations. Our results suggest that if the activation of MBCs is due to the 
sublimation of water-ice, the ice must be buried in the top few meters. We note that the
uncertainties due to the SPH method are approximately 0.5 -- 0.9 m.

\subsection{Implications for the area of ice exposure}

\subsubsection{Interior Surface of the Impact Crater}

An inspection of figures 3 and 4 indicates that in addition to the bottom of a crater, ice may also be 
exposed on the crater's interior surface. This suggests that although during an impact, some of the 
target material, including its ice is lost due to vaporization and scattering to the space and regions 
outside the crater, a considerable amount of ice may still remain on the inner surface of the crater and 
contribute to the activation of the MBC. To calculate the interior area of an impact crater, we fit the 
crater with  a spheroid and use the fitted spheroid to determine the crater's diameter. Knowing the 
crater depth as explained in the previous section, we calculate the interior area of a crater by 
calculating the surface area of the fitted spheroid. Figure 7 shows the results for all our head-on 
collisions. As expected, the interior area of an impact crater increases with increasing the target's 
water content. That is due to the fact that, as shown by the columns on material moduli in Table 2,
while basalt shows much resistance to deformation,
water-ice is significantly less resistant to compression, shear, and ductile deformation/yield. 
As a result, a material being a blend of basalt and water-ice will resist an impact to a 
lesser extent if its water-ice fraction becomes higher. This in turn translates into larger craters 
with increasing water content. 

Figure 7 shows that despite the fact that tuff is even weaker than water-ice with respect to deformation after an impact,
adding ice inclusions into tuff also results in larger craters. The reason can be attributed to tuff being a non-brittle, 
ductile material that maintains its ductile state and does not become cohesionless and liquid-like when it gets damaged by an impact.
As shown in figure 7, this effect gets more pronounced for higher impact energies that result in higher degrees 
of damage to the water ice.

Assuming that the water-ice is present under a thin layer of surface material, the exposed water-ice 
on the crater walls of an MBC will have significant contribution to its activity. 
Table~\ref{t:mbcactdata} lists the surface areas of the active regions of the currently known MBCs. 
The footnote at the bottom of this table compares the recent data with other observation-based 
estimates to illustrate an expected order-of-magnitude certainty-level. 
Our simulations yield crater surfaces of up to a few hundred square meters which, compared to the values in
Table~\ref{t:mbcactdata}, are too small, by 1-2 orders of magnitude,  
to account for the observed activation. If activity is to be explained by impact 
craters, these results suggest that multiple impact sites may be active and/or the water content of
the object may be larger than the anticipated 5\,\% so that crater sizes are larger and expose more 
water-ice. Also, as explained in the next section, possible re-accreted 
ejected material can provide an additional mechanism for supplying sublimating water-ice after an impact.

\subsubsection{Area of Ice-Exposure on the surface of the MBC}

The area of ice-exposure is not limited to only the bottom and interior surface of an impact crater.
Parts of the material ejected out of a crater may be re-accreted on the surface of the target and
form an area where ice may be exposed. The amount of ejected material, its water content, and the amount of 
re-accretion strongly depend on the impact velocity as well as the material strength and gravity 
of the target. To determine the mass of the re-accreted ice and its exposed area, we followed the dynamical evolution of the 
scattered ejecta by identifying fragments via a friend-of-friends algorithm with the SPH 
smoothing length as the discriminating criterion. Our analysis indicated that, because of the low escape velocities 
of objects with sizes similar to those of the currently known MBCs, most of the fragments that are generated in high 
velocity impacts are scattered out of the system. Only when the impact velocities are low and the target has a 
high water content, some of the fragments will have low enough velocities to be re-accreted. 
Figure 8 shows the amount of the ejected mass in terms of the velocities of the ejecta for impact 
velocities between 100\,m ${\rm s}^{-1}$ and 1000\,m ${\rm s}^{-1}$ and an impact angle of $30^\circ$. 
The impactor was m-sized. The target was taken to be a mix of basalt and ice with 0\% and 50\%
water content. For the sake of comparison, a pure-ice target was also considered. 
As shown here, most of the scattered material is ejected with velocities larger than 5\,m ${\rm s}^{-1}$.
This velocity is a few m ${\rm s}^{-1}$ larger than the escape velocity of a typical MBC which is smaller 
than 2.5\,m ${\rm s}^{-1}$  (see Table 1). Considering 2.5\,m ${\rm s}^{-1}$ as the uppermost limit of an MBC's 
escape velocity, figure 8 indicates that only a small amount of the ejected material may be re-accreted by the 
target. It is expected that slower impacts onto targets with higher water contents yield more re-accreted material.

To demonstrate the implications of these results, we assume, merely for the sake of argument, that MBC 
133P/(7968) Elst-Pizzaro is made of a mix of basalt and water-ice, and has been subject to the impact 
scenario of figure 8. Figure 9 shows the amount of the re-accreted mass on the surface of this (hypothetical) 
MBC, after an impact, in terms of the impact velocity and for two values of water-mass ratio of 
0 and 50\%. The case of a pure-ice MBC is also shown for a comparison. As shown here, an increase of the impact 
velocity from 100\,m ${\rm s}^{-1}$ to 500\,m ${\rm s}^{-1}$ decreases the amount of re-accreted material by an 
order of magnitude. While an accreted mass of $10^4$ kg for an impact velocity of 100\,m ${\rm s}^{-1}$, combined 
with the exposed ice at the bottom and interior area of the impact crater may be enough to ignite the activity 
of 133P/(7968) Elst-Pizzaro, other MBCs such as 259P/Garradd (P/2008 R1) and P/2010 R2 (La Sagra) that have 
smaller escape velocities will not be able to accrete as much material. The fact that these and other MBCs 
do show activities suggests that MBCs may contain more water than considered here and/or their activities may 
be due to ice-sublimation in several active craters resulted from multiple impacts at low velocities.

\section{Summary and Concluding Remarks}

We carried out an extensive study of the impact of a m-sized body with a km-sized 
object to examine the viability of the impact scenario as a mechanism to expose embedded water-ice 
and trigger the activation of main belt comets. We simulated impact events considering 
different water-mass fractions for very hard and very soft targets. We showed that ice can be exposed on the 
bottom and interior surface of an impact crater and may also be re-accreted on the surface of an MBC,
creating a large combined area of ice sublimation. Results of our simulations indicated that

\begin{itemize}

\item considering the current range of impact velocities in the asteroid belt, the depth of an impact crater
is only slightly larger than 10 m even when the target is assumed to be of the softest material. This suggests 
that if the activation of MBCs is due to the sublimation of water-ice, ice must be buried in the top few meters
of the object. The latter is consistent with the model by \citet{Schorghofer08} who showed that a layer of dust with a
thickness of only a few tens of cm can prevent water-ice in asteroids from sublimating for the age of the solar system;

\item substantial amount of ice may be exposed on the inner surface of an impact crater contributing to the
activation of an MBC. The size and therefore the inner area of a crater is larger for targets with higher
water-mass fraction; 

\item some of the ejected material may be re-accreted when the impact velocities are low (e.g., $\lesssim$ 100 m/s),
and may contribute to the activation of MBCs. Calculations by \citet{Hsieh04} and \citet{Hsieh06} suggested a very 
large ice-exposure area to account for the activity of MBC 133P/(7968) Elst-Pizzaro. Our simulations 
indicated that while for an object similar to this MBC, the exposed ices on the bottom and interior surface
of an impact crater, combined with the ice re-accreted on the surface of the object may be large enough to
account for its activation, in general, due to the small gravity of MBCs, the amount of re-accreted material on
the surfaces of these objects is too small to contribute to their activity. The latter implies that the activity of MBCs
is most likely due to ice-sublimation from multiple impact craters and/or the water content of these bodies are 
larger than the anticipated 5\%, providing more ice in exposed areas for sublimation.

\end{itemize}

Although the simulations presented here portray the impact scenario as a viable mechanism for exposing sub-surface ice
to ignite the  activity of MBCs, they contain some limiting assumptions. First, we did not consider
porosity for our targets. Including porosity is a complicated task that requires developing complex mathematical models. 
The latter is currently under development and we expect the results to be ready for publication in not so distant future. 
However, given the simulations presented here, we do not expect 
the result to change drastically even when porosity is included. 

We considered the target material to be basalt or tuff. However, MBCs are most likely made of 
carbonaceous chondrites. That means, in order to simulate impact events, the knowledge of the
equation of state of this material as well as the flaw distribution for the fracture model are required, which are unknown.
Our approach, that is, simulating impacts using a very hard (basalt) and a very soft (tuff) target allows us
to portray a general picture of the impact scenario as a mechanism for ice exposure that can be applied to a variety 
of targets with material strengths between these two ranges.

We believe that our SPH model of elasto-plastic continuum mechanics with brittle failure for applicable 
materials covers a valid range of target substances, even though for most water-ice weight fractions, 
the bulk densities of our materials (see Table 2) differ from known densities of MBCs of about 
$1300\,\mathrm{kg\,m^{-3}}\/$ to $1400\,\mathrm{kg\,m^{-3}}\/$ 
\citep{Hsieh04,Hsieh06,Jewitt09,hsimee11,hsi14,Hsieh15}. Also, our damage and plasticity models 
account for reduced material strength and cohesion when under load, for instance by an impact, 
down to entirely cohesionless material at the impact site or during a break-up \citep{hirsch14}.

Our simulations did not account for the vaporization of ice inclusions due to the heat of an impact. It is expected that 
after an impact, some of the embedded ice will vaporize and leave the target. Although some of this vaporized ice may
also condense and settle on the surface of the target, when put in the context of the activity of the currently known
MBCs, the loss of ice due to vaporization implies that a single impact 
on a small, km-sized MBC may not be able to expose sufficient amount of ice to account for its activity. The latter
suggests that the activity of MBCs is likely due to ice sublimation from multiple impact sites, and/or
the water content of MBCs (and other asteroids in the outer part of the asteroid belt, 
for that matter) may be higher than the conventional 5\% that is traditionally considered in models of 
terrestrial planet formation to provide more ice in the exposed areas.

Finally, we would like to note that we did not consider a regolith layer on the top of a target's icy material. 
We assumed ice to exist everywhere and
considered a random distribution of icy inclusions throughout the target. While it is expected that the inclusion of a 
regolith layer would decrease the depth of an impact crater, the micro-impacts created by the scattered fragments 
of this layer may shatter the surface of the target in other regions to the extent that the underlying ice may be exposed 
on a larger area. The latter is currently being studied.

\acknowledgments

We would like to thank the referee for critically reading our paper and for his/her valuable comments that
improved our manuscript. NH acknowledges support from NASA PAST program under grant NNX14AJ38G.
TIM and RD acknowledge support from FWF Austrian Science Funds under projects S11603-N16.

\clearpage
\begin{figure}
\includegraphics[scale=0.6]{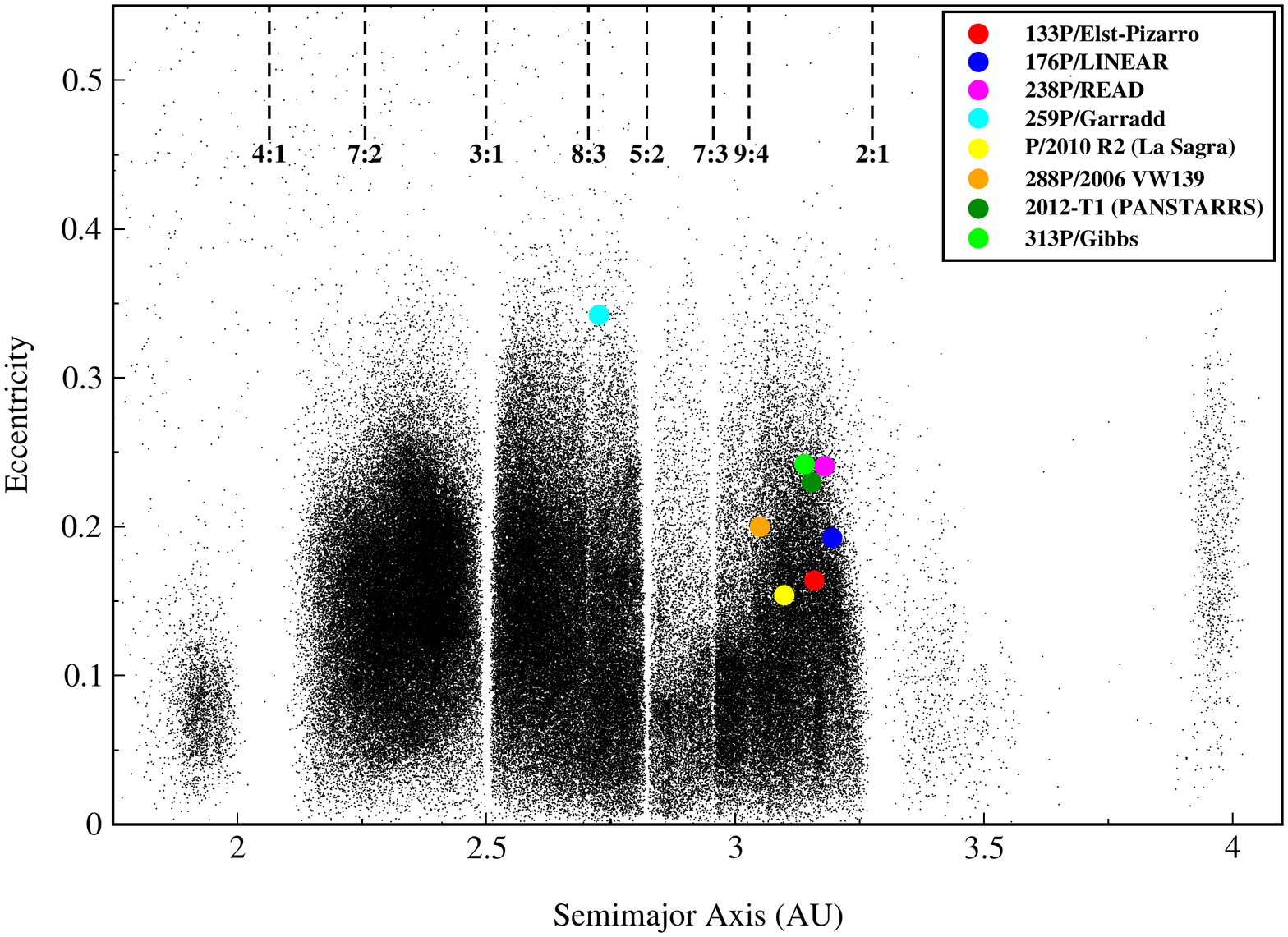}
\caption{Locations of the currently known MBCs in the asteroid belt. The background shows all asteroids and the positions of
 mean-motion resonances with Jupiter.}
\end{figure}

\clearpage
\begin{figure}
\includegraphics[scale=0.6]{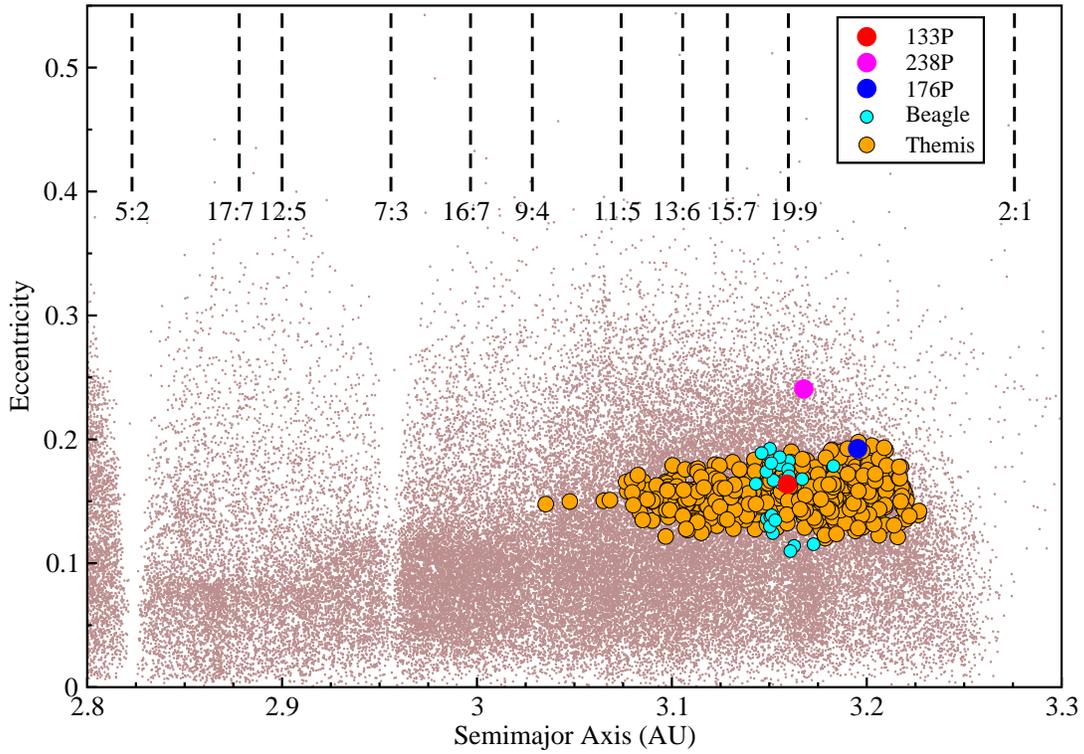}
\caption{Association of MBCs 133P/(7968) Elst-Pizzaro, 176P/(118401) LINEAR, and 238P/READ (P/2005 U1)
with the Themis and Beagle families. The background shows a portion of the asteroid belt that is in the vicinity of 
the Themis family and the locations of mean-motion resonances
with Jupiter.}
\end{figure}

\clearpage
\begin{figure}
\center
\includegraphics[scale=0.9]{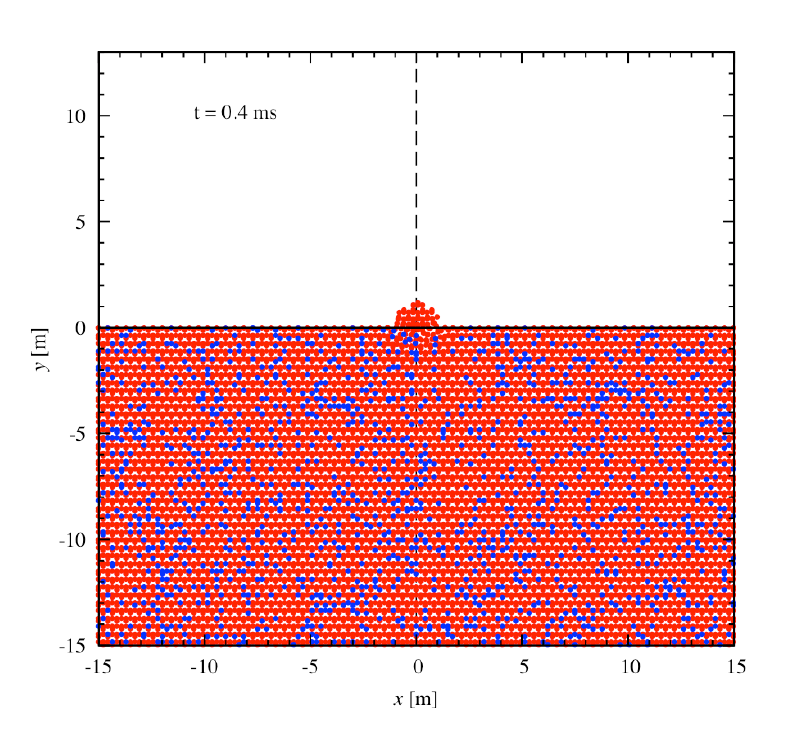}
\includegraphics[scale=0.9]{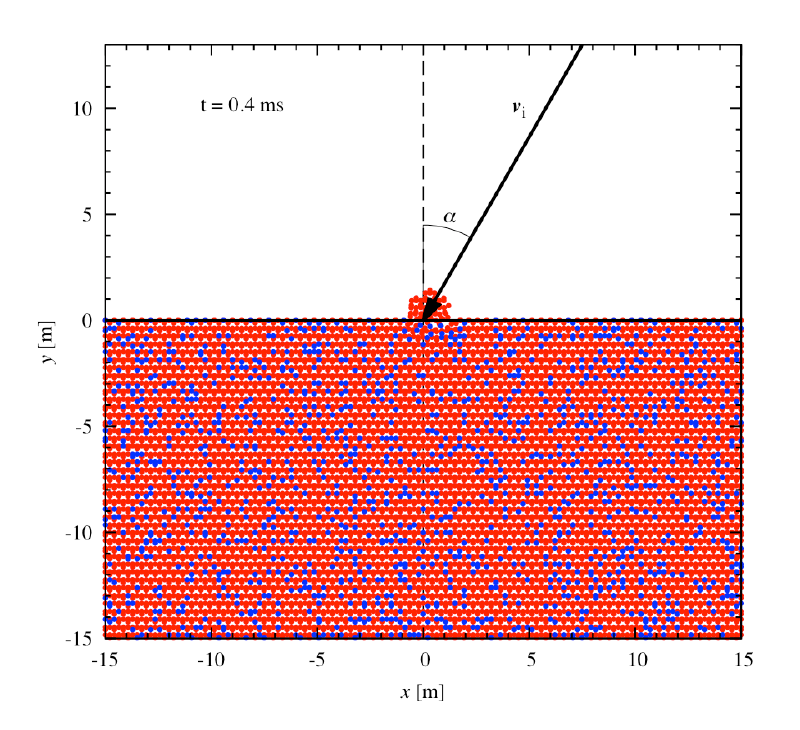}
\vskip -5pt
\includegraphics[scale=0.9]{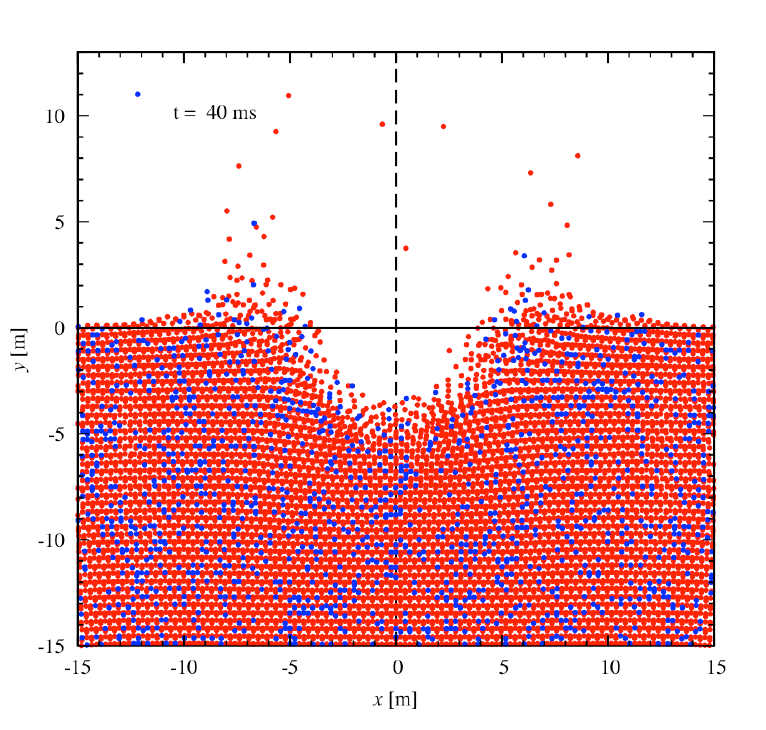}
\includegraphics[scale=0.9]{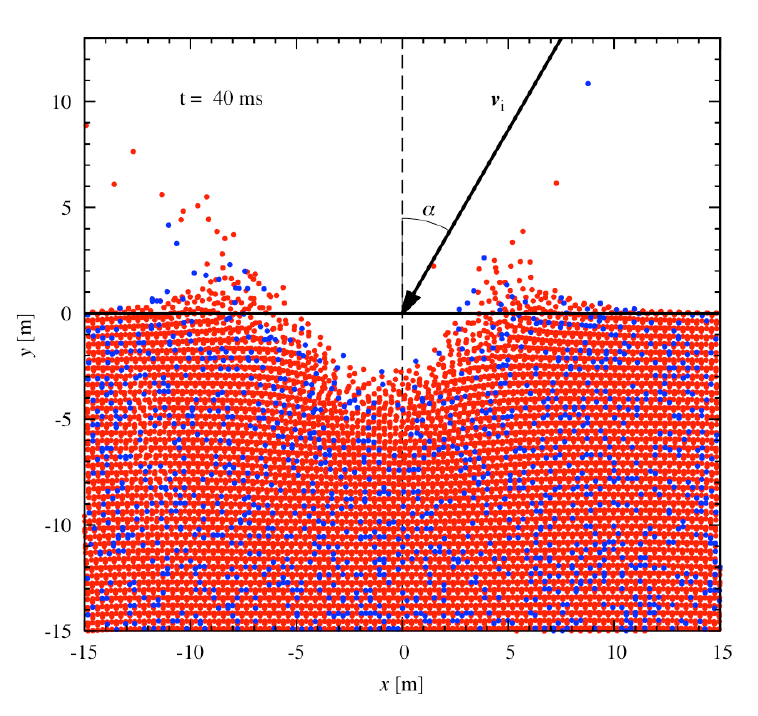}
\vskip -5pt
\includegraphics[scale=0.9]{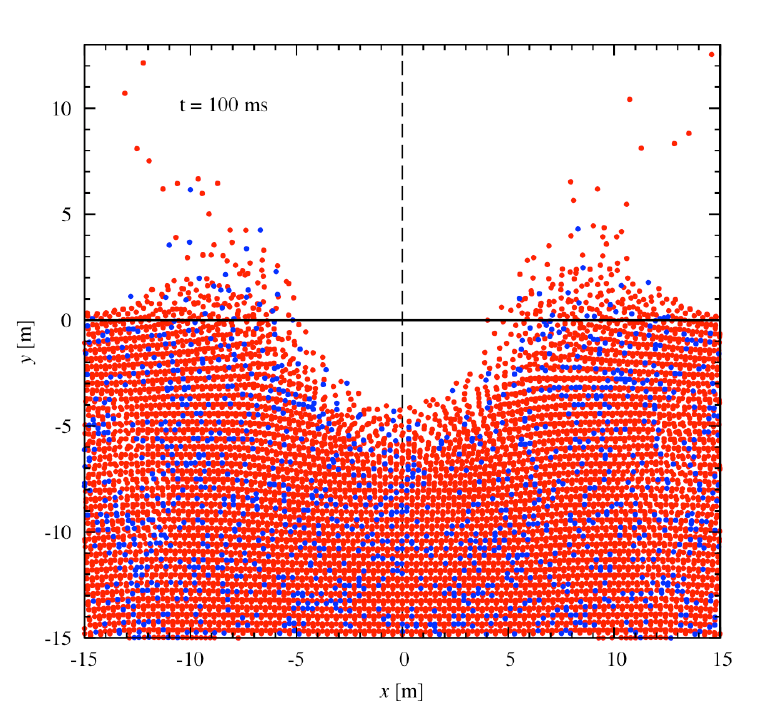}
\includegraphics[scale=0.9]{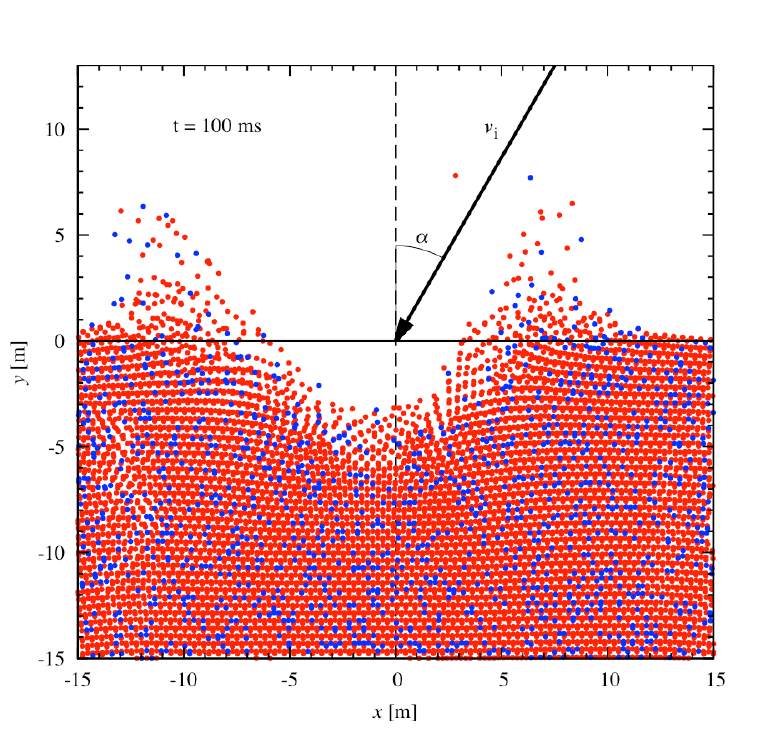}
\vskip -10pt
\caption{Snapshots of the collision of a m-sized object with a basaltic target with a 5\% water-mass fraction.
The impactor is made of pure basalt. The impact velocity is 4.4\,km ${\rm s}^{-1}$ and $\alpha=0$ (left) 
and $30^\circ$ (right). As shown here, after the impact, ice is exposed at the bottom and on the interior 
surface of the crater.}
\end{figure}

\clearpage
\begin{figure}
\center
\includegraphics[scale=0.9]{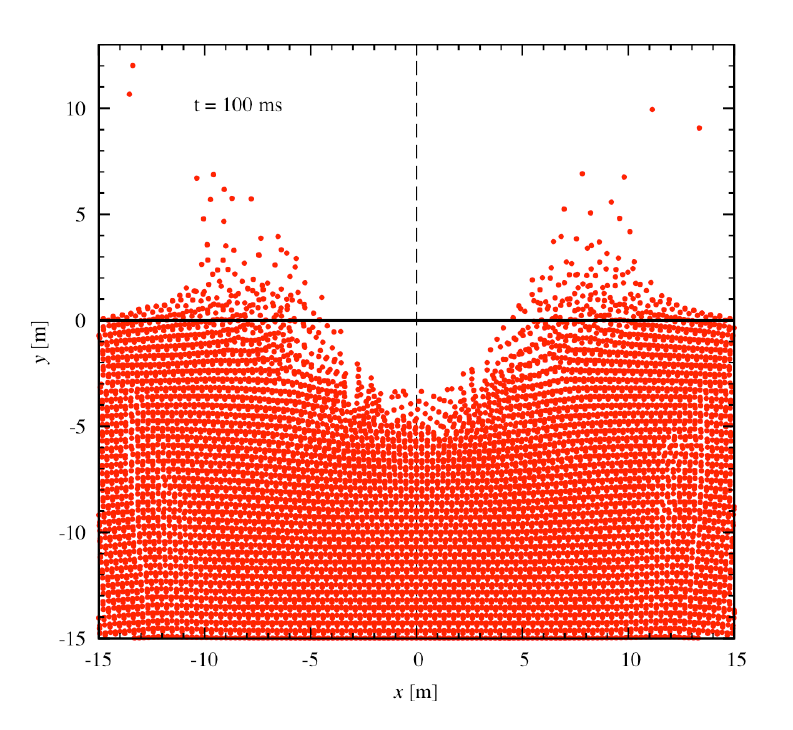}
\includegraphics[scale=0.9]{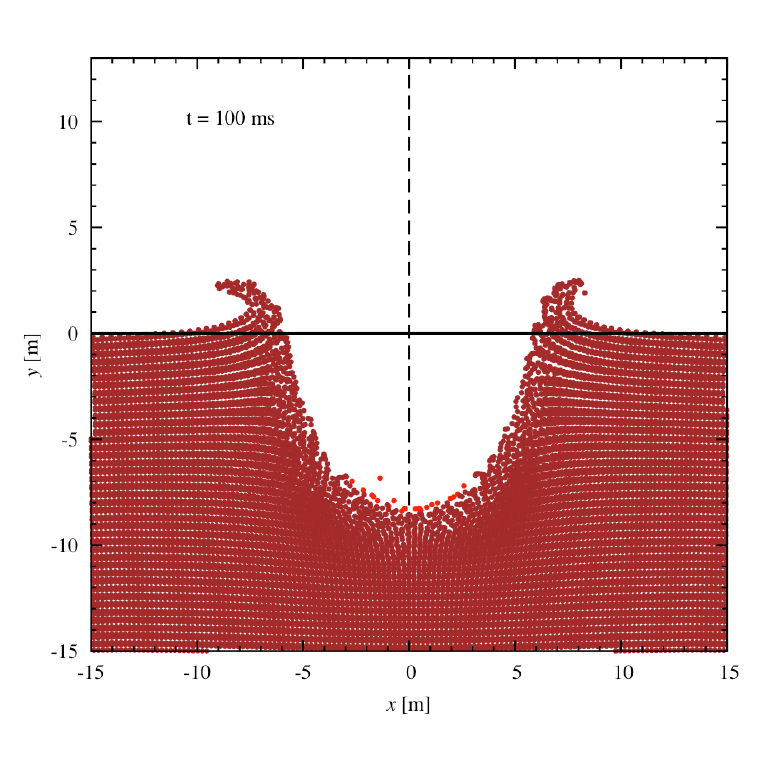}
\vskip -5pt
\includegraphics[scale=0.9]{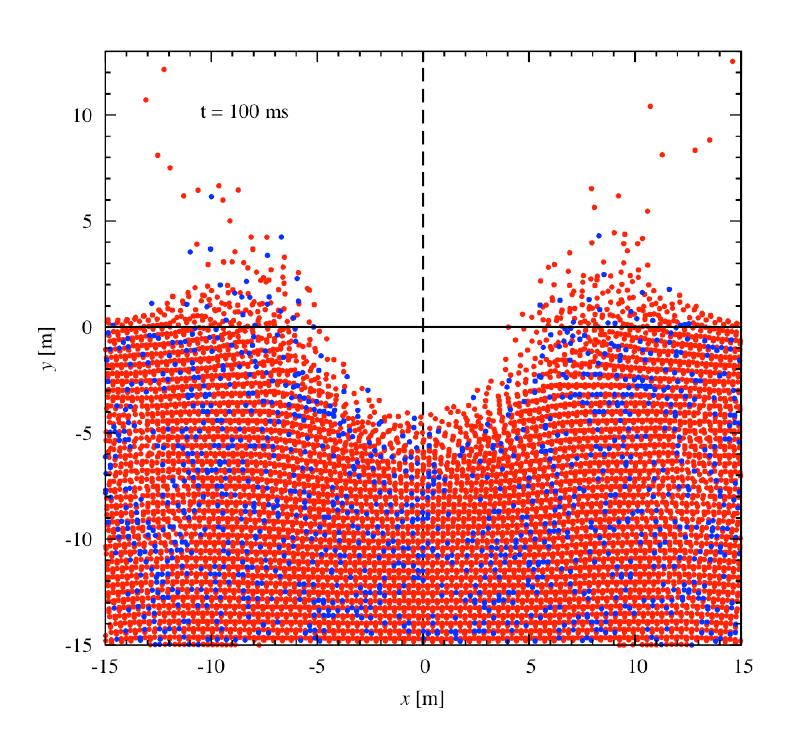}
\includegraphics[scale=0.9]{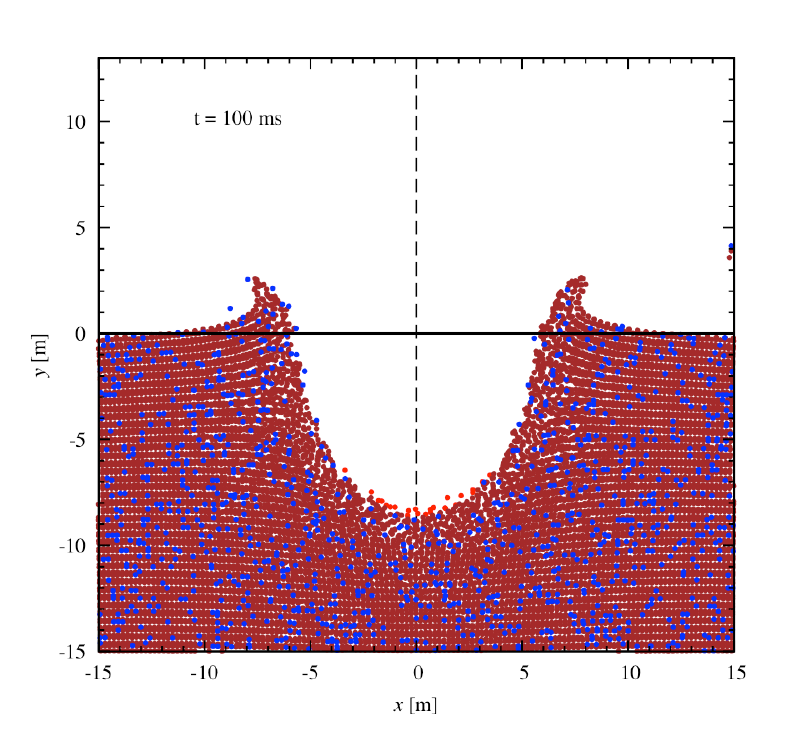}
\vskip -5pt
\includegraphics[scale=0.9]{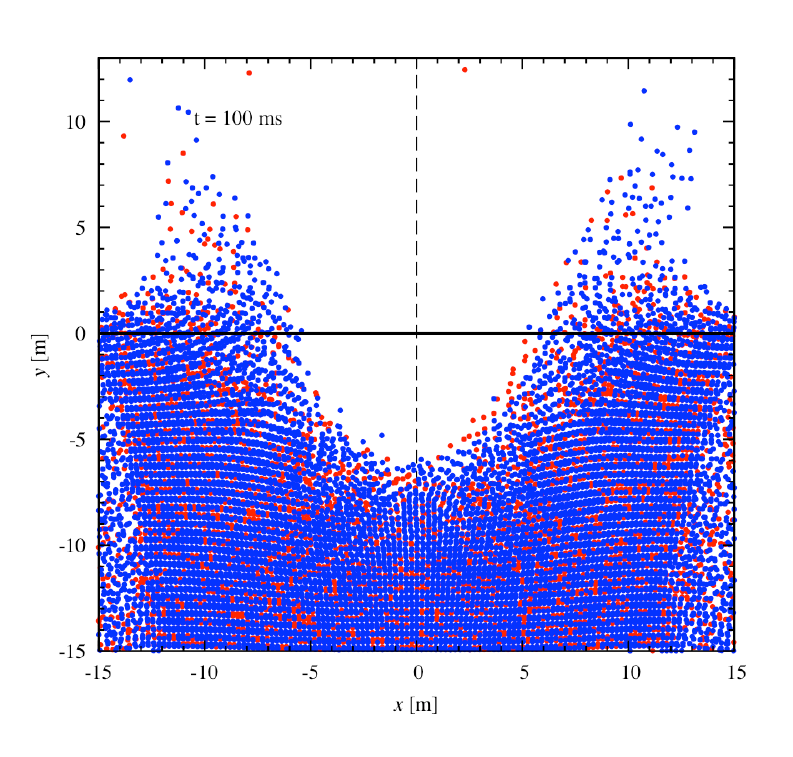}
\includegraphics[scale=0.9]{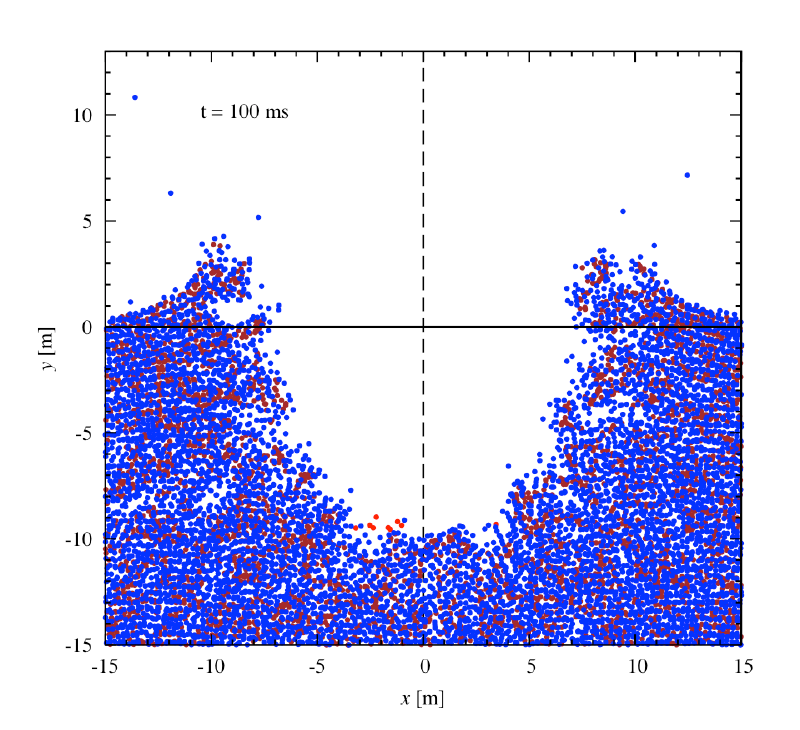}
\vskip -10pt
\caption{State of the impact crater in a 4.4\,km ${\rm s}^{-1}$ head-on collision (as in figure 3) 
100 ms after the impact. The target on the left column is made of basalt and the one on the right is made of tuff. 
From top to bottom, the water-mass fraction of the target is 0, 5\% and 50\%, respectively.}
\end{figure}

\clearpage
\begin{figure}
\center
\includegraphics[scale=1]{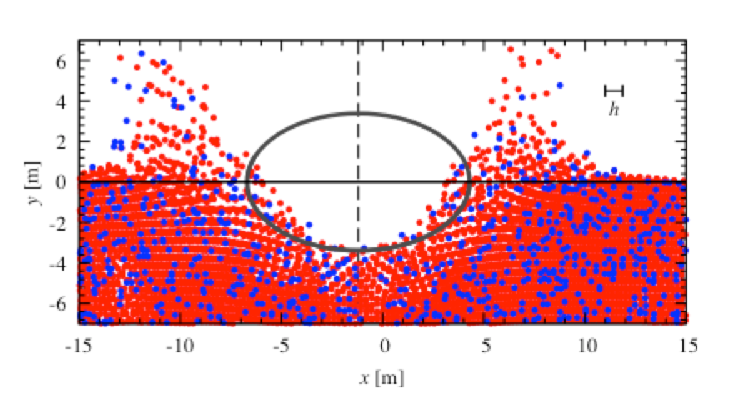}
\includegraphics[scale=1]{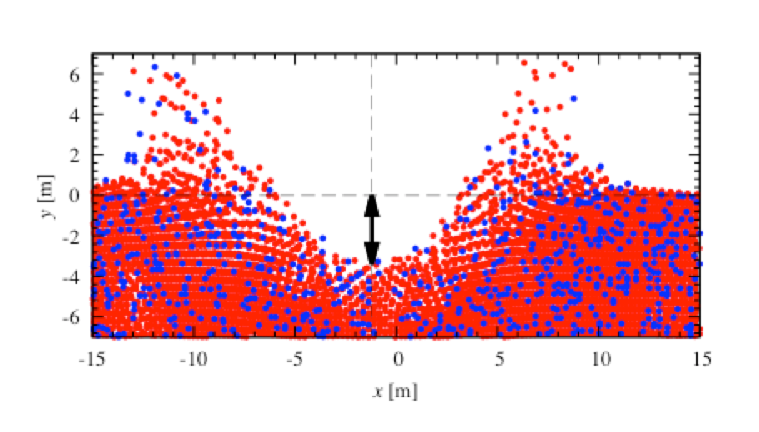}
\vskip -10pt
\caption{Determining the depth of an impact crater. The target is made of basalt with a 5\% water-mass fraction.
The impact velocity is 4.4\,km ${\rm s}^{-1}$ and the impact angle is $30^\circ$. The panel on the left shows the fitted ellipse 
in the the cross section of the crater along the vertical component of the impact velocity (the $x$-component).
The fitted ellipse is slightly extended along the $z$-axis, perpendicular to the page. The panel on the right shows
the depth of the crater after identifying the lowest point of the fitted elliptical band at 3.39 m.}
\end{figure}

\clearpage
\begin{figure}
\center
\includegraphics[]{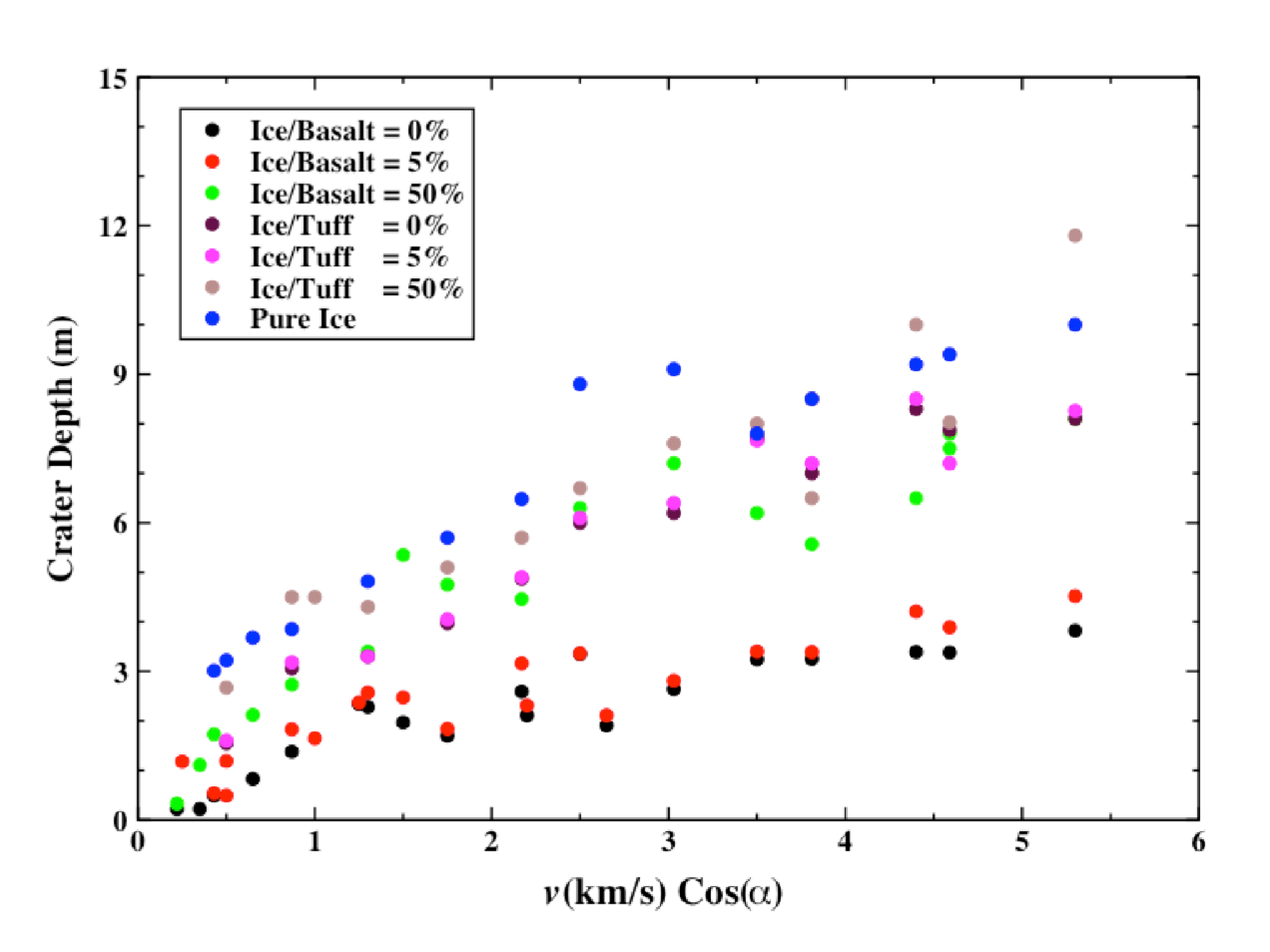}

\vskip 10pt
\caption{Variations of crater depth in terms of the magnitude of the vertical component of the impact velocity
and the water content of the target for basalt, tuff and pure ice. As shown here, the depth of the crater increases by increasing
the target's water-mass fraction. However, it does not exceed 12 m.}
\end{figure}

\clearpage
\begin{figure}
\center
\includegraphics[]{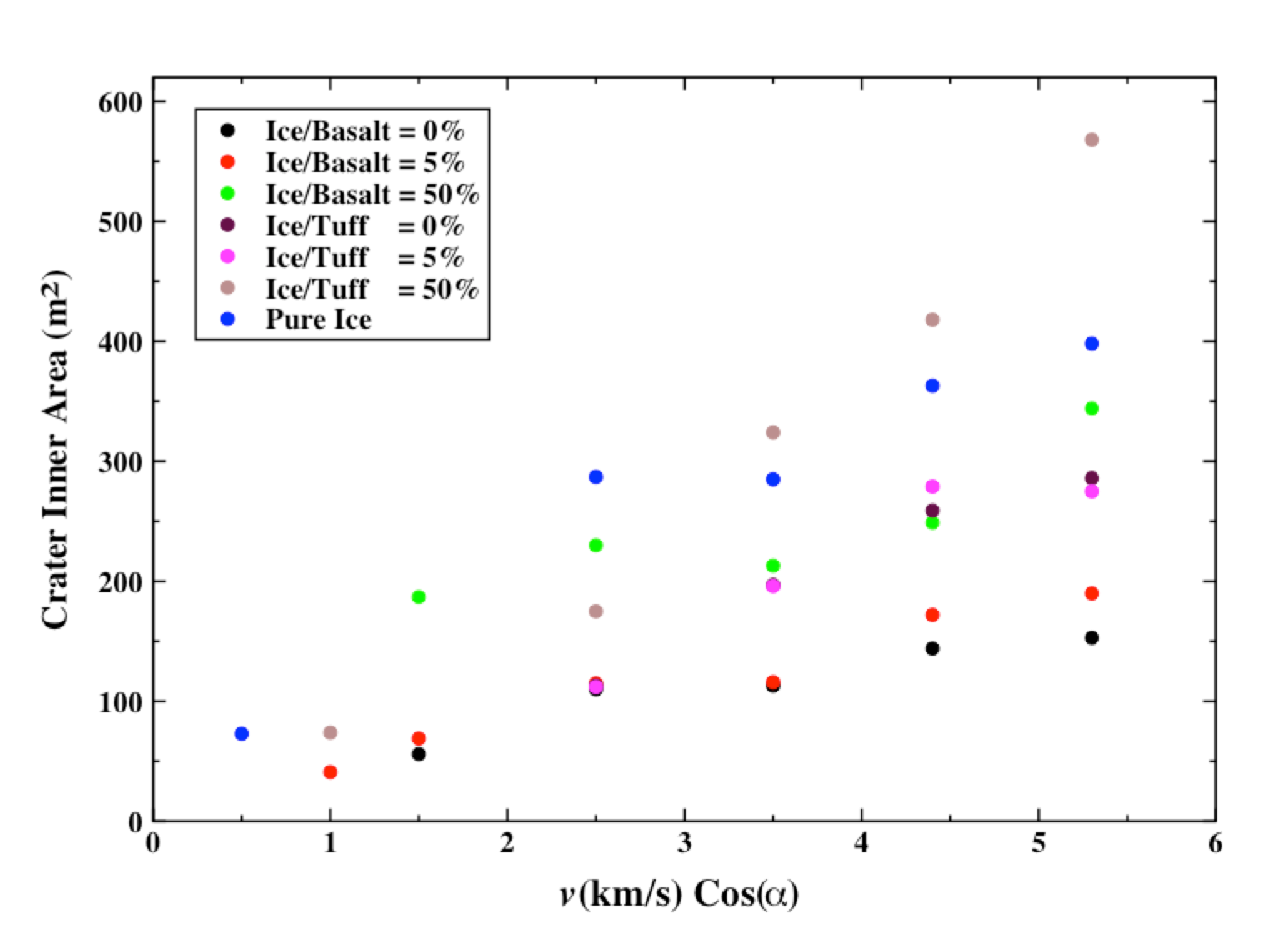}
\vskip 10pt
\caption{Variations of the inner area of an impact crater in terms of the impact velocity in a head-on collision.
The target is made of basalt or tuff with different water-mass fractions, or it is pure ice. As shown here, the 
inner area of a crater increases by increasing the target's water content.}
\end{figure}

\clearpage

\begin{figure}
\center
\vskip -1in
{\includegraphics[scale=1.2]{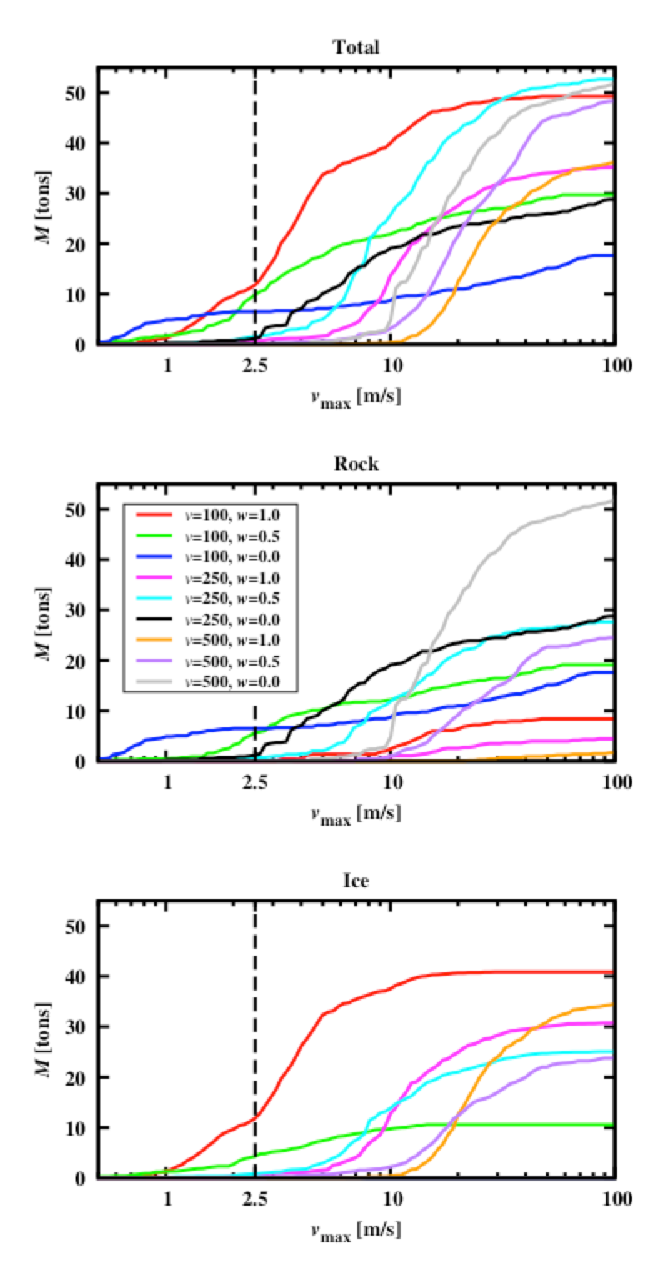}
\includegraphics[scale=1.2]{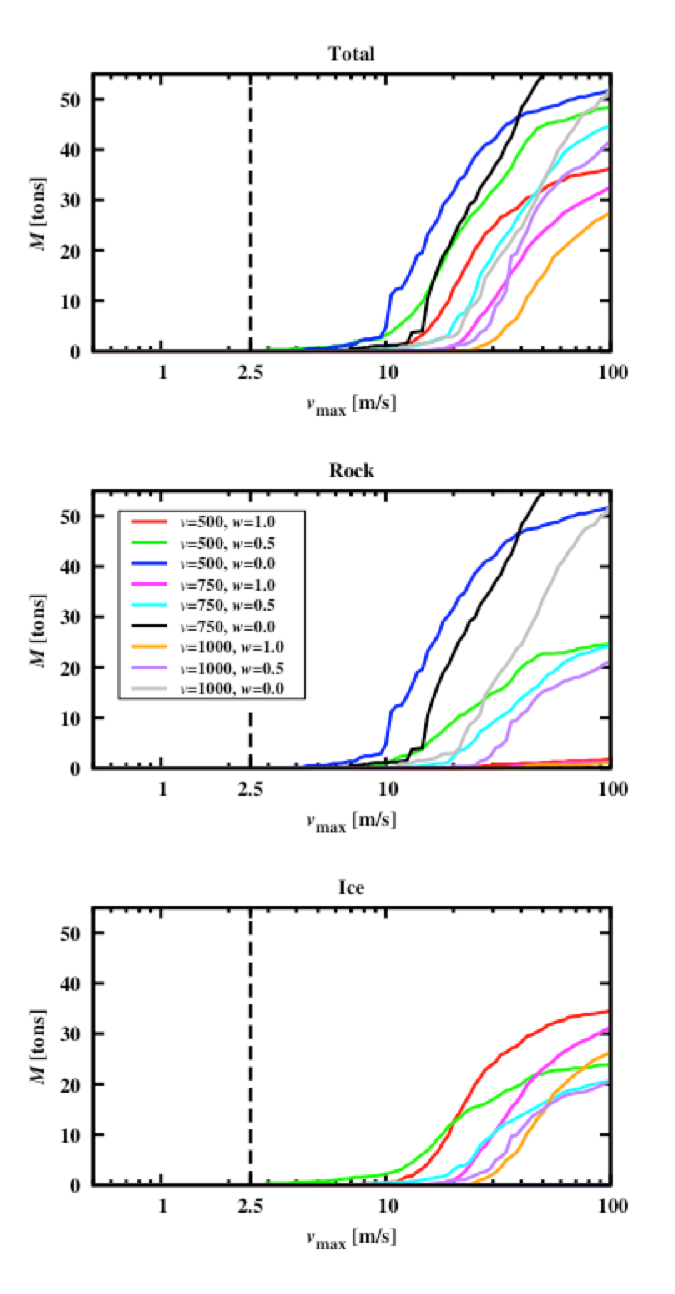}}
\caption{Graph of the ejected mass $(M)$ in terms of the maximum ejection velocity  $({v_{\rm max}})$
for low, intermediate, and high impact velocities. The impactor is m-sized and the impact angle is $30^\circ$. 
The target is made of basalt with water-mass fraction of $w =0$ or 0.5, or is pure ice with $w=1.0$. The
dashed line at 2.5\,m ${\rm s}^{-1}$ represents the upper limit of escape velocities of the currently known MBCs.}
\label{f:cummvsv}
\end{figure}

\clearpage

\begin{figure}
\center
{\includegraphics[scale=0.5]{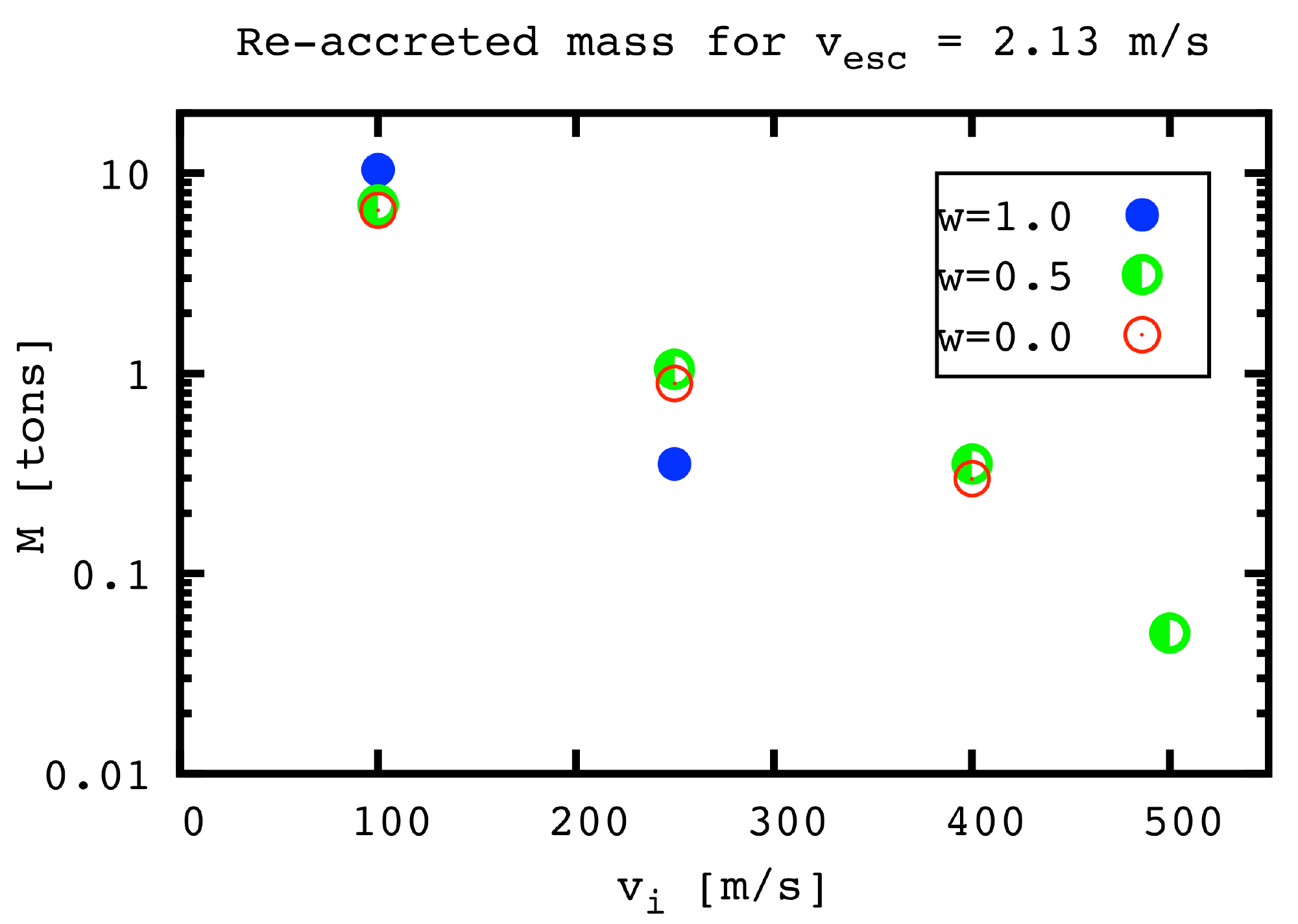}}
\caption{Re-accreted mass by a hypothetical MBC with the same size as that of 133P/(7968) Elst-Pizarro
($v_\mathrm{esc}= 2.13$\,m ${\rm s}^{-1}$) but made of basalt 
in terms of the target water content $w\/$ and impact velocity $v_\mathrm{i}\/$ at a $30^\circ$ angle.
\label{f:macc}}
\end{figure}

\clearpage

\begin{deluxetable}{lcccccc}
\tablecaption{Physical properties of the currently known MBCs\label{t:mbcdata}. The quantity $D_{\rm e}$ represents
an MBC's effective diameter, $a\,, e$ and $i$ are its semimajor axis, eccentricity and orbital inclination,
respectively, $T_{\rm J}$ is its Tisserand number with respect to Jupiter, and $v_{\rm esc}$ is the value of its
escape velocity. Adopted from \citet{Jewitt15}.}
\tablehead{
\colhead{Object}  & \colhead{$D_\mathrm{e}$ [km]} & \colhead{$a\, \mathrm{[AU]}$} & \colhead{$e$}  &
\colhead{$i \mathrm{[deg]}$}   & \colhead{$T_{\rm J}$}  &  \colhead{$v_\mathrm{esc}\, \mathrm{[m/s]}\/$}}
\startdata
133P/(7968) Elst-Pizarro  &   $3.8\pm0.6$	&   3.157	&  0.165   &  1.39  &  3.184   &  2.13   \\
176P/(118401)LINEAR	  &   $4.0\pm0.4$	&   3.196	&  0.192   &  0.24  &  3.167   &  1.95   \\
238P/Read (P/2005 U1)     &   0.8               &   3.165	&  0.253   &  1.27  &  3.152   &  ....   \\
259P/Garradd (P/2008 R1)  &   $0.3\pm0.02$      &   2.726	&  0.342   &  15.90 &  3.216   &  0.62   \\
P/2010 R2 (La Sagra)	  &   1.1       	&   3.099       &  0.154   &  21.39 &  3.098   &  0.49   \\
288P/(300163) 2006 ${\rm VW}_{139}$  & 3        &   3.050       &  0.200   &  3.24  &  3.203   &  ...    \\
P/2012 T1 (PANSTARRS)     &   2.4               &   3.154       &  0.236   &  11.06 &  3.134   &  ...    \\
313P/Gibbs (P/2014 S4)    &   1.0	        &   3.156       &  0.242   &  10.97 &  3.132   &  0.86   \\
\enddata
\end{deluxetable}

\begin{deluxetable}{lccccccccccccc}
\tabletypesize{\scriptsize}
\rotate
\tablecaption{Material parameters for basalt, ice, and tuff. The quantity $\varrho_0$ is the bulk density of the object.
The 10 quantities ${\rho_0},\,{A_{\rm T}},\,{B_{\rm T}},\,{E_0},\, {E_{\rm {iv}}},\,{E_{\rm {cv}}},\, {a_{\rm T}},\, 
{b_{\rm T}},\, {\alpha_{\rm T}}$ and 
$\beta_{\rm T}$ are the parameters used in the Tillotson equation of state \citep{Melosh96}. The remaining quantities,
$K,\, \mu$, and $Y_0$ are the bulk modulus, shear modulus, and yield stress, respectively.
Values for basalt and ice are taken from \citep{Benz99}. Note that $A_\mathrm{T}\/$ 
and $B_\mathrm{T}\/$ are set equal to the bulk modulus. For tuff, the equation of state parameters were adopted 
from \cite{Melosh96}, and $K\/$, $\mu\/$, and $Y_0\/$ are derived from measurements by \cite{Heap14} and \cite{sto69}.}
\tablewidth{0pt}
\tablehead{
\colhead{Material} & \colhead{$\varrho_0$}[$\mathrm{kg/m^{3}}$] & \colhead{$A_\mathrm{T}$ [GPa]} & 
\colhead{$B_\mathrm{T}$ [GPa]} & \colhead{$E_0$ [$\mathrm{MJ/kg}$]} & \colhead{$E_\mathrm{iv}$ [$\mathrm{MJ/kg}$]} & 
\colhead{$E_\mathrm{cv}$ [$\mathrm{MJ/kg}$]} & \colhead{{$a_\mathrm{T}$}} & \colhead{$b_\mathrm{T}$} & 
\colhead{$\alpha_\mathrm{T}$} & \colhead{$\beta_\mathrm{T}$} & \colhead{$K$ [GPa]} & \colhead{$\mu$ [GPa]} & 
\colhead{\tstrut $Y_0$ [GPa]}}
\startdata
Basalt & 2700 & 26.7  & 26.7  & 487 & 4.72 & 18.2  & 0.5 & 1.50 &  5 &  5 & 26.7 & 22.7 & 3.5 \\
Ice & 917 & 9.47 & 9.47 &10 & 0.773 & 3.04 & 0.3 & 0.1 & 10 & 5 & 9.47 & 2.8 & 1 \\
Tuff & 1700 & 4.5 & 3 & 6 & 3.5 & 18 & 0.5 & 1.3 & 5 & 5 & 4.6 & 2 & 0.013\\
\enddata
\end{deluxetable}

\begin{deluxetable}{lllcrr}
\tablewidth{0pt}
\tablecaption{The rate of the loss of water-ice at perihelion $(\dot{m}_\mathrm{w})$
and the activation area ($A_\mathrm{act}$) of the currently known MBCs calculated
using observational data from \citet{hsi14}.\label{t:mbcactdata}}
\tablehead{
\colhead{Object}           & \colhead{$\dot{m}_\mathrm{w}\,\mathrm{[kg\,s^{-1}]}\/$}      &
\colhead{$A_\mathrm{act}\,\mathrm{[m^2]}\/$}  }
\startdata
133P/(7968) Elst-Pizarro	&$7.2\times 10^{-6}$	&$2\times 10^{4}$	\\
176P/(118401) LINEAR		&$8.3\times 10^{-6}$	&$1\times 10^{3}$	\\
238P/Read (P/2005 U1)		&$1.3\times 10^{-5}$	&$2\times 10^{3}$	\\
259P/Garradd (P/2008 R1)	&$3.4\times 10^{-5}$	& - \\
288P/(300163) P/2006 $\mathrm{VW_{139}}$	&$1.1\times 10^{-5}$	&$5\times 10^{3}$	\\
P/2010 R2 (La Sagra)		&$7.6\times 10^{-6}$	&$5\times 10^{4}$	\\
P/2012 T1 (PANSTARRS)		&$1.1\times 10^{-5}$	&$1\times 10^{4}$	\\
P/2013 R3					&$1.7\times 10^{-5}$	& -	\\
\enddata
\tablecomments{\citet{Hsieh04} revised estimated $A_\mathrm{act}$ to be a few tens to a few hundreds 
of m$^2$. \citet{hsimee11} estimated $A_\mathrm{act}\approx 2\times 10^4\,\mathrm{m^2}\/$ for 238P/Read (P/2005 U1).}
\end{deluxetable}

\end{document}